\documentclass[twocolumn,showpacs,preprintnumbers,amsmath,amssymb,superscriptaddress]{revtex4}
\usepackage{graphicx}
\begin{document}

\title{Prospect for room temperature tunneling
anisotropic magnetoresistance effect: density of states anisotropies in CoPt systems}

\author{A. B. Shick}
\author{F. M\'aca}
\author{J. Ma\v{s}ek}

\affiliation{Institute of Physics ASCR, Na Slovance 2, 182 21 Praha 8, Czech Republic}

\author{T. Jungwirth}
\affiliation{Institute of Physics ASCR, Cukrovarnick\'a 10, 162 53
Praha 6, Czech Republic}
\affiliation{School of Physics and Astronomy, University of Nottingham,
University Park, Nottingham NG7 2RD, UK}

\begin{abstract}
Tunneling anisotropic magnetoresistance (TAMR) effect, discovered
recently in (Ga,Mn)As ferromagnetic semiconductors, arises from
spin-orbit coupling and reflects the dependence of the tunneling
density of states in a ferromagnetic layer on orientation of the
magnetic moment.  Based on ab initio relativistic calculations
of the anisotropy in the density of states we predict
sizable TAMR effects in room-temperature metallic
ferromagnets. This opens prospect for
new spintronic devices with a simpler geometry as these do not require
antiferromagnetically coupled contacts on either side of the tunnel
junction. We focus on several model systems ranging from simple
hcp-Co to more complex ferromagnetic structures
with enhanced spin-orbit coupling, namely bulk and thin film
L1$_0$-CoPt ordered alloys and a monatomic-Co chain at a Pt
surface step edge.
Reliability of the predicted density of states anisotropies is
confirmed by comparing quantitatively our  ab initio results
for the magnetocrystalline anisotropies in these systems with
experimental data.
\end{abstract}
\maketitle

The area of condensed matter research that aims at exploiting
synergies of magnetic and semiconducting properties in solid state
systems has served as an important test bed for understanding basic
physics and discovering new applications in spintronics
\cite{Dietl:2003_a,Zutic:2004_a}. The anomalous Hall effect and the
tunneling anisotropic magnetoresistance (TAMR) studies are two
examples of the research that provided main motivation for the work
presented here. The former study demonstrated
\cite{Jungwirth:2002_a,Onoda:2002_a,Jungwirth:2003_b,Dietl:2003_c,Yao:2004_a}
that a successful theory of the anomalous Hall effect in (Ga,Mn)As
ferromagnetic semiconductors, based on spin-orbit (SO) coupling
effects present in the band structure of Bloch states, can be
directly applied to conventional metallic ferromagnets such as Fe,
and describe quantitatively this fundamental transport coefficient. The TAMR effect
\cite{Gould:2004_a} is an offspring of attempts to develop hybrid
metal/semiconductor spin-valve devices which
revealed that a spin-valve-like response can be achieved without the
seemingly fundamental switching sequence between parallel and
antiparallel magnetizations in two ferromagnetic contacts with
different coercivities. Instead, a single ferromagnetic material
with SO interaction is sufficient for realizing the sensing or
memory functionality through TAMR whose phenomenology is even richer
than that of conventional giant magnetoresistance or tunneling
magnetoresistance effects. For example, both lower and higher
resistance states can be obtained at saturation depending on the
external magnetic field orientation, i.e., the TAMR device can act
as a sensor of the absolute direction of the external field
\cite{Ruester:2004_a}.

The TAMR was discovered in a (Ga,Mn)As/AlO$_x$/Au stack
\cite{Gould:2004_a} and confirmed by subsequent experiments in
(Ga,Mn)As based vertical \cite{Ruester:2004_a} and planar
\cite{Giddings:2004_a} tunnel devices. The former experiment
underlined the importance of high quality interfaces and barrier
materials for the magnitude of the effect. The lithographically
defined planar nanodevice allowed to demonstrate a direct link
between the TAMR in the tunneling nanoconstriction and normal
anisotropic magnetoresistance  in the ferromagnetic lead. Since the latter effect
also originates from SO coupling  and has been observed
in many metallic systems the work suggests that the TAMR  has
been overlooked in conventional room temperature ferromagnets. To
explore this possibility we follow a theoretical
strategy applied previously to (Ga,Mn)As TAMR devices which is based
on calculating tunneling density of states (DOS) anisotropies in the
ferromagnet with respect to the orientation of the magnetic moment
and which assumes a proportionality between the DOS and tunneling
conductance anisotropies.

The paper is organized as follows: We start with a simple
hcp-Co crystal as a bench mark material and then add to the
complexities of the studied structures in order to enhance SO
coupling related effects. The next system we explore is bulk
L1$_0$-CoPt alloy \cite{yermakov,11}. Here Co produces large
exchange splitting resulting in the Curie temperature of 750~K while
the heavy elements of Pt  substantially increase the strength of SO
coupling in the band structure of the alloy. Effects of reduced
dimensionality are explored in a thin L1$_0$-CoPt film and, reaching
the ultimate nanoscale regime, in the monatomic-Co chain at  Pt
surface step edge \cite{gambardella,shick03,ujfallusi03}. An
important part of the presented work is a simultaneous analysis of
magnetocrystalline anisotropies in the studied structures. Given the
predictive nature of our theoretical conclusions for tunneling DOS
anisotropies, calculations of a physical quantity originating from
the same SO coupled ab initio band structure  and directly
comparable to existing experimental data is particularly valuable.
Magnetocrystalline anisotropies, and the derived magnetization
reversals in external magnetic fields, are also important
characteristics that define functionality of a TAMR device.

We use the relativistic version of the full-potential linearized
augmented plane-wave method (FP-LAPW) \cite{singh}, in which SO
coupling is included in a self-consistent second-variational
procedure \cite{shick97}. The conventional (von Barth-Hedin) local
spin-density approximation  is adopted, which is expected to be
valid for itinerant metallic systems. The magnetic force theorem
\cite{force} is used to evaluate the DOS anisotropy and the
magnetocrystalline anisotropy energy (MAE): starting from
self-consistent charge and spin densities calculated for the
magnetic moment $M_S$ aligned along one of principal axes, the $M_S$
is rotated and a single energy band calculation is performed for the
new orientation of $M_S$. The DOS and magnetocrystalline
anisotropies result from SO coupling induced changes in the band
eigenvalues. Importantly, the same set of k-points has to be used
for the integration over the Brillouin zone (BZ) for accurate
evaluation of the DOS anisotropy and of the MAE. Furthermore, in
order to increase the accuracy in DOS evaluation, the smooth Fourier
interpolation scheme of Pickett {\em et al.} \cite{warren88} is used
together with linear tetrahedron method \cite{lehman}.

\begin{figure}[h]
\vspace*{-0.4cm}
\includegraphics[width=3.2in]{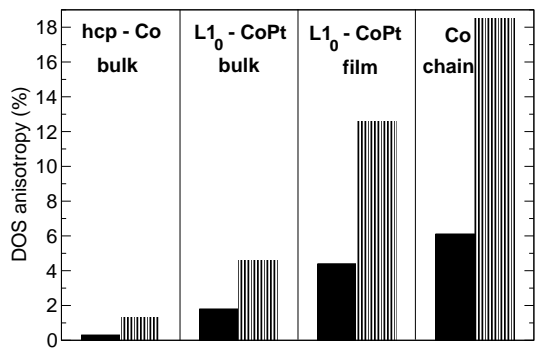}
\vspace*{-1cm} \caption{
Bulk systems: integral (filled bars) and tunneling (half-filled bars) DOS anisotropy,
$|N_{I(T)}(x)-N_{I(T)}(z)|/{\rm min}(N_{I(T)}(x),N_{I(T)}(z))$.
L1$_0$-CoPt thin film: integral (filled bar) and Pt-surface  (half-filled bar) DOS anisotropy.
Monoatomic Co chain: $|N_{I}(x)-N_{I}(z)|/{\rm min}(N_{I}(x),N_{I}(z))$ for A-step
(filled bar) and $|N_{I}(x)-N_{I}(y)|/{\rm min}(N_{I}(x),N_{I}(y))$ for B-step
(half-filled bar). 
} 
\label{ados}
\end{figure}

We start by investigating  the possibility of TAMR in elemental
transitional metal ferromagnets. Bulk hcp-Co is a theoretically well
understood uniaxial ferromagnet with the MAE of 60~$\mu$eV per unit
cell and the easy axis of magnetization along the $z$-[0001]
crystallographic direction \cite{LB}. To evaluate the DOS
anisotropy, we first performed the self-consistent FP-LAPW
calculation of the band structure  of hcp-Co using experimental
lattice constant values and  fixing $M_S$ along the $z$-[0001] axis.
The DOS anisotropy is obtained by rotating $M_S$ from the $z$-[0001]
axis to the in-plane $x$-[1000] axis. 1200~$k$-points were used for
the BZ integration. The integral DOSs at the Fermi energy for $M_S$
along the $x$-[1000] axis, $N_I(x)$, and  $z$-[0001] axis, $N_I(z)$,
are given in Tab.~1. Corresponding DOS anisotropy defined as,
$|N_I(x)-N_I(z)|/{\rm min}(N_I(x),N_I(z))$ is plotted in
Fig.~\ref{ados} (see black bars). The integral DOS anisotropy $\sim
0.3$\% is relatively weak, similar to the weak magnetocrystalline
anisotropy, which is a result of a small value of SO coupling in
elemental 3$d$-metals. Assuming high crystalline quality of an
hcp-Co based TAMR device with a large degree of in-plane momentum
conservation during the tunneling we can improve our estimate of the
effect by considering the tunneling DOS at the Fermi energy
\cite{igor99},
$N_T  = 1/\Omega_{BZ}\;
\int_{BZ} d^3{k} {v_z}^2(E_F) \delta (E(\vec{k})-E_F)$,
where $v_z=\partial E(\vec{k})/\partial k_z$ is the group velocity
component along the tunneling $z$-direction and $E_F$ is the Fermi
energy.  The corresponding anisotropy (see blue bar in
Fig.~\ref{ados}) is a factor of 4 larger than the anisotropy in
$N_I$ which is a trend seen previously in the (Ga,Mn)As
\cite{Gould:2004_a}. The 1.3 \% anisotropy in $N_T$ suggests a
measurable, albeit weak, TAMR effect even with a simple hcp-Co.

A natural recipe for enhancing the TAMR effect is by using ordered
mixed crystals of magnetic 3$d$ and non-magnetic 5$d$ transition metals. 
In these systems the
magnetic atoms can produce large exchange fields polarizing the
neighboring heavy elements with strong SO coupling. Pt is
particularly favorable because of its large magnetic susceptibility.
Here we consider the L1$_0$-CoPt alloy which is known to have a
strong MAE with the easy-axis aligned perpendicular to alternating
Co and Pt metal layers \cite{yermakov}. The large anisotropy results
from broken cubic symmetry in this layered alloy and from the
presence of heavy elements of Pt, as explained above. Our previous
relativistic FP-LAPW calculations \cite{11} yield the MAE of
1.03~meV, in a very good agreement with experimental data
\cite{yermakov}. DOS anisotropy calculations performed here follow
the same numerical procedures as in Ref.~\cite{11}. Starting from
self-consistent calculations for $M_S$ aligned along the $z$-[001]
axis, we rotate $M_S$ to the $x$-[110] direction and calculate the
corresponding DOS. 952~$k$-points in the irreducible part of BZ
(3584~$k$-points in full BZ) are used in these numerical
simulations. As expected, both the  $N_I$ anisotropy of 1.8\% and
$N_T$ anisotropy of 4.6\% (see Tab.~1 and Fig.~\ref{ados}) in this
mixed crystal are substantially larger than their hcp-Co
counterparts.
\begin{table}[h]
{\footnotesize
\begin{tabular}{ccccccc}
\hline
\multicolumn{1}{c}{hcp-Co} &\multicolumn{1}{c}{$\vec{M}_S
\parallel x$ axis} & \multicolumn{1}{c}{$\vec{M}_S \parallel  z$
axis} \\
$N_I$ &     1.999 &             2.004   \\
$N_T$ &   6.780 &             6.696    \\
 \hline
\multicolumn{1}{c}{L1$_0$-CoPt-bulk} &\multicolumn{1}{c}{$\vec{M}_S
\parallel x$ axis}&\multicolumn{1}{c}{$\vec{M}_S \parallel  z$
axis} \\
$N_I$&   2.091     &   2.055\\
$N_T$ & 10.709    & 11.205\\
\hline \multicolumn{1}{c}{L1$_0$-CoPt-film}
&\multicolumn{1}{c}{$\vec{M}_S
\parallel x$ axis}&\multicolumn{1}{c}{$\vec{M}_S \parallel  z$
axis} \\
$N_I$ &   13.310     & 12.745 \\
\hline
\multicolumn{1}{c}{} &  \; $N_I^{\uparrow} \; + \; N_I^{\downarrow}
\; \; \; = \; \; \; N_I^{tot}$ &\; $N_I^{\uparrow} \; + \;
N_I^{\downarrow} \; \; \; \;  = \; \;
N_I^{tot}$ \\
Pt-Surface   & 0.489 + 0.485 = 0.974 & 0.385 + 0.480 = 0.865\\
Co-subsurface & 0.231 + 1.637 = 1.868 & 0.187 + 1.615 = 1.802\\
\hline
\end{tabular}
} \caption{DOS in hcp-Co and L1$_0$ CoPt: Total integral DOS $N_I$
(1/eV) and tunneling DOS $N_T$ (eV(a.u.)$^2$) DOS at $E_F$ for
$\vec{M}_s \parallel x$ and $z$ axes. 
For
L1$_0$-CoPt film we show also layer- and spin- projected DOS.
The $k$-space convergence
achieved in the numerical calculations, using the smooth Fourier interpolation scheme \cite{warren88},
gives error-bars $< 1$\% for $N_I(x)$ in hcp-Co, $<2.5$\% for $N_T$ in  hcp-Co, and
$<0.5$\% for $N_I(x)$ and $N_T$ in L1$_0$ CoPt.}
\label{table1}
\end{table}

So far we have demonstrated prospects of room temperature TAMR and
confirmed expected trends in the TAMR with increasing strength of SO
coupling in well understood conventional ferromagnets. In these bulk
systems, ab initio relativistic calculations are known to provide
accurate description of their properties, including the subtle, SO
coupling related magnetocrystalline anisotropy. The two model
systems discussed in the following paragraphs allow us to take a
different view at the DOS anisotropy which might be more closely
related to the TAMR in real tunneling structures and, in the latter
case, explores metallic TAMR in the ultimate nanoscale limit.
Ferromagnetic electrodes in magnetic tunnel junctions are grown in
thin films in which surfaces/interfaces effects are known to play an
important role. To analyze these effects in our context  we evaluate
DOS anisotropies of a free standing L1$_0$-CoPt film consisting of 5
Co and 6 Pt layers with the same interlayer distances as in the
bulk. 400~$k$-points in the two-dimensional BZ were used in the
numerical calculations. The resulting $N_I$ anisotropy of the film
is more than a factor of 2 larger than in the bulk L1$_0$-CoPt.
Assuming a vacuum tunnel barrier and realizing that tunneling
transport characteristics are dominated by properties of electronic
states near the barrier, we can use the calculated DOS spatially
resolved to individual layers  to make a refined estimate of the
TAMR effect in the L1$_0$-CoPt film. The partial DOS corresponding
to surface Pt  and subsurface Co monolayers are listed in Tab~1. The
anisotropy in the DOS of the surface Pt layer is close to 13\%. We
note that the MAE per two-dimensional unit cell is about factor of 5
larger than the anisotropy per three-dimensional unit cell in the
bulk L1$_0$-CoPt. Recalling that our CoPt film has 5 Co and 6 Pt
layers the MAE in the two structures are very similar. Since Curie
temperatures in the film and bulk L1$_0$-CoPt can also be expected
to be similar ($\sim 750$~K) the magnetic tunnel junction based  on
thin film L1$_0$-CoPt is a good candidate for observing the TAMR
effect at room temperature.

Recently, Gambardella {\em et al.} \cite{gambardella} reported
ferromagnetism below 15~K in  monatomic-Co chains at the Pt(997)
surface step edge.  While this system is unlikely to lead to room
temperature spintronic applications, it provides a unique
opportunity to study both theoretically and experimentally magnetic
and transport anisotropies in monatomic ferromagnetic chains. The
experiments \cite{gambardella} revealed  an unexpectedly strong MAE
($\sim 2.0 \pm 0.2$ meV/Co atom) with the easy magnetization axis
directed along a peculiar angle of 43$^{\circ}$ towards the Pt step
edge and normal to the Co chain.

\begin{figure}[h]
\includegraphics[angle=-90,width=3.4in]{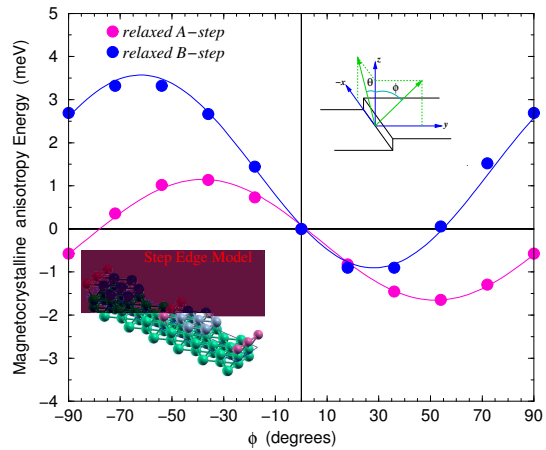}
\caption{Monatomic-Co chain: MAE as a function of the magnetization angle
$\phi$. Solid lines are  fits to the
numerical points [$1.145 \; - \; 2.801 \cos^2(\phi - 52^{\circ})$]
eV (\textit{A-step}) and [$3.530 \; - \; 4.474 \cos^2(\phi -
28^{\circ})$] eV (\textit{B-step}) . Left inset: schematic
crystal structure, used to represent the Co chain at the Pt(111)
surface B-step edge. Right inset: polar angles
$\phi$ and $\theta$.} 
\label{cowire}
\end{figure}

We analyzed the experiment by considering two possible Pt(111) surface step edge
geometries, the $\langle 100\rangle$ microfaceted A-step and the 
$\langle 111\rangle$ microfaceted B-step \cite{sheffler}.
In the left inset of Fig.~\ref{cowire} we show model supercells for the B-step
which consist of a sub-subsurface
and a subsurface Pt layer built of 6 rows of Pt atoms, while the
surface step is modeled by 3 rows of Pt, one Co row, and two rows of
empty Pt sites. Vacuum was modeled by the equivalent of two empty Pt
layers. All interatomic distances in the $y-z$-plane were relaxed
using scalar-relativistic atomic forces \cite{optimization}. This
represents an important improvement over previous calculations
\cite{shick03, ujfallusi03} which assumed values for pure Pt.

The self-consistent relativistic FP-LAPW calculations were performed
for magnetization aligned along the $z$-axis, using 180~$k$-points
in a quasi-2D BZ. The MAE is shown in Fig.~\ref{cowire}. As already
described in Ref. \cite{shick03}, the step-edge removes any
high-symmetry directions in the $y-z$-plane. The easy axis is tilted
from the $z$-axis towards the Pt step edge by 52$^{\circ}$ for the A-step
and by 28$^{\circ}$ for the B-step, in 
good agreement with the experimentally observed \cite{gambardella}
angle of 43$^{\circ}$. The total
energy difference between states with the magnetization along the
hard and easy axes is $\approx$ 2.8 meV/Co (A-step) and 4.5 meV/Co
(B-step); the experimental value of $\sim$ 2 meV/Co was obtained from
magnetic reversal measurements at temperature 45~K. Higher
experimental MAE and therefore even better quantitative agreement
between our calculations and experiment is expected at temperatures below the
ferromagnetic transition temperature in this system.
\begin{table}[h]
{\footnotesize
\begin{tabular}{ccccc}
 \hline
&\multicolumn{1}{c}{$\vec{M}_S \parallel x$ axis}&
\multicolumn{1}{c}{$\vec{M}_S \parallel y$
axis}&\multicolumn{1}{c}{$\vec{M}_S \parallel z$ axis}\\
\hline {\mbox{\textit{A-step :}}}
$E_{x,y} \; - \; E_z$ & 8.54        &  -0.571             \\
$N_I(E_F)$  &   26.225      & 24.963        & 24.719      \\
\hline {\mbox{\textit{B-step :}}}
$E_{x,y} \; - \; E_z$ &-0.190       &   2.639             \\
$N_I(E_F)$  &   24.731      & 29.313        & 26.606      \\
\hline
\textit{Anisotropy constants} & $K_1$       &  $K_2$        &  $K_3$  \\
\hline
relaxed \textit{A-step}  & -1.99       &   2.28        &  1.16   \\
relaxed \textit{B-step}  & -0.63      &  -0.71       & -2.00   \\
unrelaxed \textit{A-step} (Ref. \cite{shick03})     &-1.70 & -0.12        & -1.34  \\
unrelaxed \textit{B-step} (Ref. \cite{ujfallusi03}) & -0.16 & -1.06 & -4.81  \\
\hline
\end{tabular}
} \caption{The MAE and DOS in monatomic-Co chain: Total energy
differences $E_x - E_z$, $E_y - E_z$ (meV),
 integral DOS ($N_I(E_F$), 1/eV) for $\vec{M}_s
\parallel x, \; y$ and $z$ axes, and
magnetocrystalline anisotropy constants (meV).}
 \label{table2}
\end{table}

The energy differences between states with $M_S$ along  the $x$-axis
(along the Co-chain) or $y$-axis (in-plane, perpendicular to the
chain), and the  $z$-axis (out-of-plane, perpendicular to the chain)
are shown in Tab.~II. For the A-step, the MAE is relatively weak in
the $y-z$-plane, and it becomes very strong when $M_S$ is rotated
towards the $x$-axis. For the B-step, the MAE is stronger in
$y-z$-plane, and becomes weaker for $x-z$-plane. Furthermore, we can
evaluate the magnetic anisotropy constants by fitting our results to
the total energy angular dependence \cite{ujfallusi03},
[$K_1 \cos2\theta+ K_2 (1 - \cos2\theta) \cos2\phi
 +  K_3 \sin2\theta \sin\phi$],
where $K_{1,2,3}$ are the uniaxial anisotropy constants, and
$\theta$ and $\phi$ are conventional polar angles. The calculated
values of the  anisotropy constants are shown in Tab.~II and
compared with previous results assuming no structural relaxation
\cite{shick03, ujfallusi03}. The Co-chain anisotropy constants,
$\sim$2 meV/Co, set a record, exceeding the bulk and surface
anisotropies for the conventional transitional metal materials.
Similarly, the anisotropy in $N_I$ is large. For the A-step, e.g., we obtained a (6.1 \%)
anisotropy in the integral DOS, indicating a sizable TAMR effect.
Note that for the one-dimensional
chain we can make an alternative estimate of the magnitude of the
TAMR based on the calculated difference between the number of 
energy bands at $E_F$ for the two $M_S$ directions. We obtained 25
bands for $M_S$ along the $z$-axis and 26 bands for $M_S$ parallel
to the $x$-axis. The corresponding 4\% effect is consistent with the
estimate based on the calculated DOS anisotropies. For the B-step,
the change in $N_I$ for different magnetization orientations is particularly large for $M_S$ along the $x$- and
$y$-axes. The corresponding DOS anisotropy is 18.5\%.

To conclude, our  theoretical results for DOS and magnetocrystalline
anisotropies in CoPt structures, based for both quantities on the
same ab initio relativistic band structures and in the latter one
agreeing quantitatively with available experimental data,  suggest
sizable TAMR effects in these metal ferromagnets.
While the anisotropies in the ferromagnetic material make the TAMR
possible, the magnitude of the effect can be very sensitive to
parameters of the entire tunnel device, most notably of the geometry
and crystalline quality of the tunnel junction, as demonstrated in
(Ga,Mn)As based devices \cite{Ruester:2004_a,Giddings:2004_a}.
The integral DOS anisotropies calculated here for
the CoPt systems are larger than their
(Ga,Mn)As counterparts.

We acknowledge fruitful discussions with B.L. Gallagher, C. Gould,
B. Gurney, K. Ito, S. Maat, A.H. MacDonald, E. Marinero, L.W.
Molenkamp, P.M. Oppeneer, W.E. Pickett, J. Sinova, and J.
Wunderlich, and support from Grant Agency of the Czech
Republic under Grant No. 202/05/0575, by Academy of Sciences of the
Czech Republic under Institutional Support No. AV0Z10100521, by the Ministry of Education of the 
Czech Republic Center for Fundamental Research LC510, and by the UK
EPSRC under Grant GR/S81407/01.

\end{document}